# Inferring brain-computational mechanisms with models of activity measurements

Nikolaus Kriegeskorte[1] & Jörn Diedrichsen[2]

[1] Medical Research Council Cognition and Brain Sciences Unit, Cambridge, UK
[2] Western University, London, Ontario, Canada

## Abstract

High-resolution functional imaging is providing increasingly rich measurements of brain activity in animals and humans. A major challenge is to leverage such data to gain insight into the brain's computational mechanisms. The first step is to define candidate brain-computational models (BCMs) that can perform the behavioural task in question. We would then like to infer, which of the candidate BCMs best accounts for measured brain-activity data. Here we describe a method that complements each BCM by a measurement model (MM), which simulates the way the brain-activity measurements reflect neuronal activity (e.g. local averaging in fMRI voxels or sparse sampling in array recordings). The resulting generative model (BCM-MM) produces simulated measurements. In order to avoid having to fit the MM to predict each individual measurement channel of the brain-activity data, we compare the measured and predicted data at the level of summary statistics. We describe a novel particular implementation of this approach, called probabilistic RSA (pRSA) with measurement models, which uses representational dissimilarity matrices (RDMs) as the summary statistics. We validate this method by simulations of fMRI measurements (locally averaging voxels) based on a deep convolutional neural network for visual object recognition. Results indicate that the way the measurements sample the activity patterns strongly affects the apparent representational dissimilarities. However, modelling of the measurement process can account for these effects and different BCMs remain distinguishable even under substantial noise. The pRSA method enables us to perform Bayesian inference on the set of BCMs and to recognise the data-generating model in each case.

## Introduction

In order to understand how the brain works, we need to build brain-computational models (BCMs) that can perform cognitive tasks, such as object recognition, planning, and motor control (Abbott 2008). If our models are to explain human brain function, they will need to perform complex computations that require rich world knowledge (Kriegeskorte 2015). A BCM that explains a feat of human intelligence, by definition, constitutes an artificial intelligence (AI) system. Unlike AI systems from engineering, however, it should also map onto the anatomical components of the brain and explain measurements of brain activity and behaviour (Yamins & DiCarlo 2016). Testing BCMs of various levels of complexity has a long history in neuroscience (for examples from neuroimaging, see Daw et al. 2006; Kriegeskorte et al. 2008b; Wittmann et al. 2008). Several studies have begun to evaluate AI-scale neural net models on the basis of brain-activity data (Yamins et al. 2013, Yamins et al. 2014, Cadieu et al. 2014, Khaligh-Razavi & Kriegeskorte 2014, Güçlü & van Gerven 2015, for reviews see Kriegeskorte 2011, Kriegeskorte 2015, Yamins & DiCarlo 2016).

On the theoretical side, much is known about the dynamical properties and computational capabilities of a wide variety of neural network models, ranging from detailed models of single biological neurons (Koch 2004) to rate-coding models of complex feedforward and recurrent networks that abstract from the biological details (Serre et al. 2007, Le Cun et al. 2015). The literature has described a rich repertoire of functions that can be performed by such networks, including feedforward categorisation, autoassociative memory, general function approximation, dynamic history compression, working memory, various forms of control, and general approximation of dynamical systems. How these building blocks work together in biological brains to achieve intelligent adaptive behaviour, however, remains largely mysterious (Eliasmith 2013; Marblestone et al. 2016). AI has recently achieved major advances using neural net models that are inspired by biological brains, but highly simplified (Le Cun et al. 2015). AI is beginning to provide the technological basis for modelling human brain computational functions in their full complexity. However, it is unclear how we should test and adjudicate among competing BCMs.

On the experimental side, technologies for brain-activity meausurement are rapidly advancing. In animals, two-photon imaging enables us to jointly measure unprecedented numbers of neurons (e.g. Ahrens et al. 2012). In humans, high-field functional magnetic resonance imaging (fMRI) enables us to image hemodynamic activity over the entire brain at resolutions approaching a cubic millimeter per voxel and in smaller volumes at submillimeter resolution (e.g. Yacoub 2008). Electrophysiological recordings at the scalp and inside the brain offer high temporal resolution and are likewise advancing to provide ever greater numbers of jointly measured channels. However, analysis is typically limited to data-driven methods and inferential contrasting of experimental stimuli. The goals of these analyses are (1) to map the brain for regions or neurons that exhibit differential activation or information about the stimuli in their distributed activity patterns and (2) to decriptively characterise the profile (known as "tuning") of each measured response across stimulus properties.

In order to bridge the gap between theory and experiment, we need to go beyond characterisation of response properties with generic statistical models. We need to use the multivariate brain-activity measurements to test BCMs that perform the information-processing tasks we are investigating.

We focus here on a particular class of experiments in sensory neuroscience, where experimental stimuli (e.g. images) are presented during measurement of brain activity. BCMs of the perceptual processing take the same stimuli as input and perform tasks such a visual object recognition. We assume that there are many measurement channels reflecting the neural population activity in a cortical area or subcortical region. The goal of the analysis is to infer which BCM best explains the measured brain activity patterns.

Methods for testing competing BCMs that have been used in previous studies are encoding models (Dumoulin & Wandell 2008, Kay et al. 2008, Mitchell et al. 2008, Naselaris et al. 2011, Gallant et al. 2012), representational similarity analysis (Kriegeskorte et al. 2008a, Kriegeskorte & Kievit 2013, Nili et al. 2014), and pattern-component modelling (Diedrichsen et al. 2011).

Encoding models predict the response of each measurement channel as a linear combination of the responses of the units of the BCM to the stimuli. In order to adjudicate among BCMs, the predictive performance of each model is estimated for a set of novel stimuli not used in fitting the encoding model. Encoding models naturally lend themselves to the mapping of tuning properties across the measurement channels, for example by using human semantic labels (rather than units of BCMs) as the predictor variables (Huth et al. 2012). However, the other techniques can also be used for mapping by applying them to single voxels or searchlights across brain locations (Kriegeskorte et al. 2006). Here we are concerned with inferring the data-generating BCM for a predefined region of interest. Encoding models can be used to adjudicate between BCMs by estimating the overall predictive performance of each BCM across all measurement channels in the brain region of interest.

Modelling each measurement channel as a linear combination of the model units might account for the way the measurement channels sample the neuronal activity patterns. In fMRI, for example, the measurement channels are voxels that reflect the activity of tens or hundreds of thousands of neurons whose activity gives rise to the hemodynamic signals washed into the voxel through the vasculature. Local averaging might be a reasonable approximate model of the measurement process. However, encoding models require fitting one weight per BCM unit for each measurement channel. This typically requires both the measurement of responses to a large training set of stimuli (in addition to the test set used to adjudicate between BCMs) and a strong prior on the weights. The particular prior used for fitting the weights is part of the hypothesis tested and can affect the results of inference on BCMs. Typically, a 0-mean Gaussian prior on the weights (i.e. L2 regularisation) is used to achieve stable fits. Complex BCMs such as deep neural networks can have hundreds of thousands of units in a single layer. While simpler models can account for certain tasks, we will ultimately need AI-scale models with many parameters to account for the complexities of the neuronal networks in biological brains. Here we explore a class of methods that do not require fitting a separate predictive model for each measurement channel. The BCM predictions are evaluated at the level of summary statistics of the activity measurements.

A summary-statistical method for testing BCMs is representational similarity analysis (RSA; Kriegeskorte et al. 2008a, Kriegeskorte & Kievit 2013, Nili et al. 2014). In RSA, the representation of each stimulus in the brain region of interest is compared to the representation of each other stimulus. This yields a representational dissimilarity matrix (RDM), which characterises the representations in terms of the stimulus distinctions it emphasises. RDMs can be computed for representations in BCMs and biological brains. Statistical inference determines which BCM best accounts for the RDM in the brain region of interest (Nili et al. 2014, Khaligh-Razavi & Kriegeskorte 2015).

An advantage of RSA is that it does not require fitting a linear model to predict each measurement channel. The RDM provides a useful characterisation of the representational geometry, that abstracts from the roles of particular responses. If the noise is isotropic (or rendered isotropic by spatial prewhitening of the responses in analysis; Walther et al. 2015) and the Euclidean distance is used as the representational dissimilarity measure, then any two representations that have the same RDM afford the same information for linear readout by other brain regions. An RDM thus arguably characterises a functional equivalence class of neuronal population codes. However, a disadvantage of direct comparison of RDMs from measurement channels and RDMs from BCM units is that the way the channels reflect the neuronal activities (e.g. as local averages in fMRI) is not accounted for. As we will see, RDMs computed from measurement channels do accurately reflect RDMs computed from the underlying neuronal activity if the measurement channels average across neurons with random weighting. However, a key observation we explore in this paper is that when a population of neurons that is spatially organised at multiple scales (as seen for many cortical areas, e.g. early visual cortex) is sampled by local averaging (as in fMRI voxels), the resulting measurement-channel RDM can differ substantially from the neuronal RDM. This needs to be taken into account when performing inference on BCMs (Carlin & Kriegeskorte 2015).

Both encoding models and RSA, in their widely used instantiations, fail to account for our knowledge and uncertainties about the measurement process. In fMRI, for example, a voxel can be approximated as a local average, suggesting the use of positive weights concentrated in a certain small patch of a BCM's internal representation corresponding to the patch of cortex sampled by the voxel. However, the prior typically used in encoding models assumes merely that the weights are small, while allowing negative weights and not constraining the spatial weight distribution. RSA so far implicitly assumed that the measurement channels sample the neurons with random weights. Although neuronal recordings and fMRI voxels can yield strikingly similar RDMs (Kiani et al. 2007; Kriegeskorte et al. 2008b), we cannot assume that the random-weight sampling assumption holds in general throughout the brain for either fMRI or neuronal recordings.

Here we argue that a generative measurement model (MM) should be integrated into statistical inference on BCMs (Fig 1). A natural solution is to simulate the effect of measurements by means of a method of sampling from the units of a BCM that mimics the way our measurements sample neuronal activity. The MM may have unknown parameters, for which prior distributions are specified. By simulating the brain computation (by the BCM) as well as the measurements (by the MM), we can predict the distribution of measurements for each BCM. In order to perform inference on a set of BCMs, we need to compare the predicted measurement distribution with the actual measurements. In order to compute the posterior over BCMs, we will estimate the likelihood for each BCM.

The method we describe has the following key novel features:

- The measured response channels are considered as *a sample from a population of response channels* that might have been measured from a given brain region.

- A *measurement model* (MM) provides a probabilistic characterisation (expressing our knowledge and uncertainties) of the process through which the measurement channels reflect neuronal activity.

- Each BCM predicts a distribution of responses across the population of potential measurement channels and the predicted distribution is compared to the data at the level of summary statistics, obviating the need for fitting a separate measurement model to each measurement channel.

- Statistical inference is performed by computing the posterior, i.e. the probability of each BCM given the data.

We use a deep neural network for visual object recognition to simulate fMRI data by taking local weighted averages. Data is simulated for the five visuotopic convolutional layers of the network, which are considered as five distinct BCMs. The simulated dataset enables us to test the proposed method for inferring BCMs, since the ground-truth computational mechanism that generated the data is known in each case. We demonstrate the effect of the measurement process on the apparent representational geometry and show that modelling the measurements, without knowledge of the precise measurement parameters (local averaging range, voxel-grid placement) enables us to infer the data-generating BCM for each of the five layers of the network.

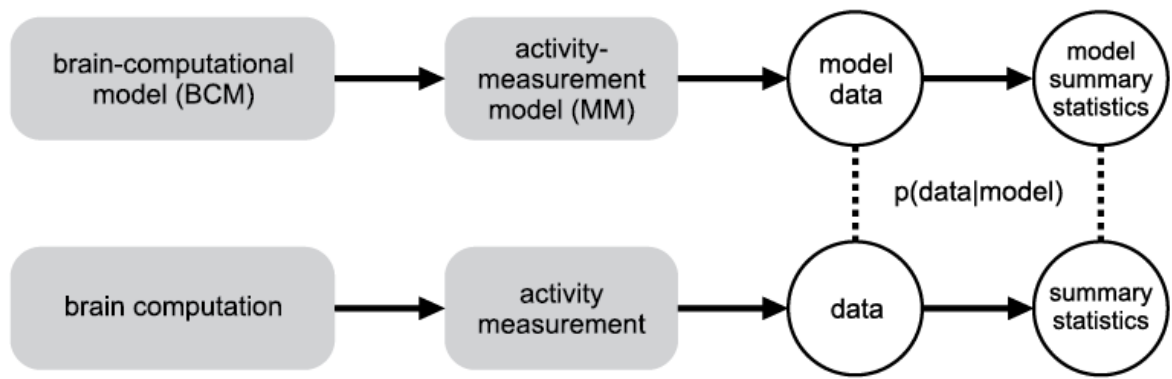

Figure 1 | **Evaluating a brain-computational model with brain-activity measurements.** To evaluate each of a number of competing BCMs, we would like to compute the likelihood p(data|model) for each model. We need to account not only for the brain computations (using a BCM) but also for the measurement process (using an MM). The MM simulates the way the measurement channels sample the units of the BCM. One approach to inference is to evaluate the likelihood at the level of the measurements. This requires fitting a parameterised MM to predict each individual measurement channel. To avoid having to fit an MM to each channel, we instead predict summary statistics of the population of possible measurement channels.

## Method

### Deep neural net model as testbed for inference on BCMs

We use the deep neural network for visual object recognition from Krizhevsky et al. (2012), known as AlexNet, as a testbed for inference on BCMs. AlexNet is a deep neural network trained by backpropagation (Werbos 1981, Rumelhart et al. 1986; for a review, see Schmidhuber 2014) to recognise, which of 1000 object categories a natural photograph displays. AlexNet uses a convolutional architecture (LeCun & Bengio 1995) inspired by the primate visual system. The first convolutional layer detects a set of features in the image. Each higher convolutional layer detects a set of features in the preceding layer. Each feature template is detected all over the two-dimensional image space by convolving the feature template with the image (or preceding layer) and passing the result through a rectifying nonlinearity (rectified linear units: negative values set to zero). As a result, the convolutional layers (first five layers) are visuotopic with receptive fields increasing from layer to layer, as in the primate visual system.

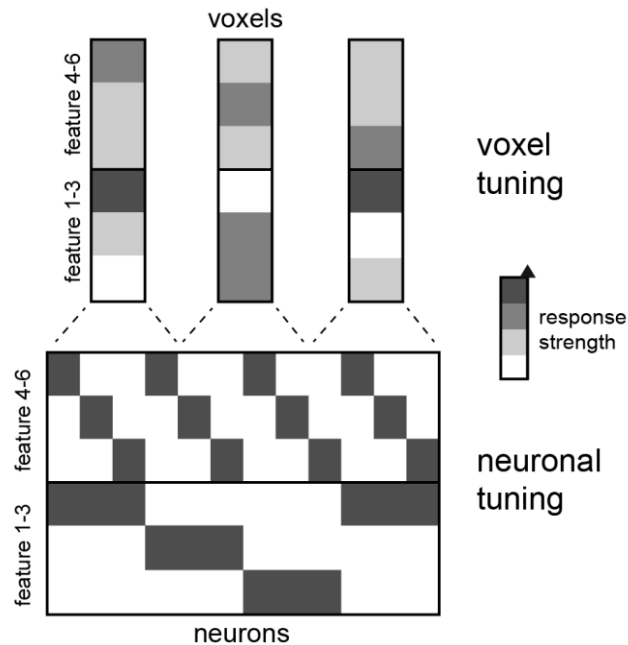

Figure 2 | **Locally averaging fMRI voxels exhibit different tuning than the neurons they sample.** A set of neurons (lined up horizontally at the bottom along a single spatial dimension) have a variety of tuning functions (columns of the bottom rectangle). The arrangement of the neurons in cortex is not random. Closeby neurons tend to have similar tuning with respect to features 1-3, but can have different tuning with respect to features 4-6. Locally averaging voxels (top) pool across neurons with different tuning. Local averaging in a voxel will attenuate neuronal selectivities that change rapidly along the cortex (features 4-6), but will approximately preserve selectivities that vary slowly along the cortex (features 1-3).

## Measurement model for BOLD fMRI

We pretend that AlexNet is a biological brain and simulate the data we would expect to obtain if we measured it with blood-oxygen-level-dependent (BOLD) fMRI (Ogawa et al. 1992). We assume that each fMRI voxel measures a local average of neuronal activity (Engel et al. 1997, Parkes et al. 2005, Shmuel et al. 2007). Voxels may reflect not only activity occurring within their boundaries, but also activity outside closeby, whose effects on on the local blood oxygen level flow into the voxel over the period of measurement. We therefore assume that each voxel signal is a Gaussian-weighted local average. The local-averaging range depends on the details of vascular physiology (including the point spread function of the vasodilatory response), the voxel size, and the cortical magnification factor, which defines what visual angle corresponds to a mm distance on the cortex in a retinotopic visual area. Since some of these parameters vary substantially (Dougherty et al. 2003) across human subjects, the precise local averaging range is unknown. We therefore assume a prior distribution over this parameter.

Each convolutional layer contains a spatial image map for each of a number of features. This corresponds to the local topological maps (e.g. of orientations in V1) nested inside the global retinotopic maps in early visual cortical areas. The fact that the model defines not only a computational process, but also the spatial arrangement of the computational units, enables us to predict how the model's internal representations would be reflected in locally averaging fMRI voxels.

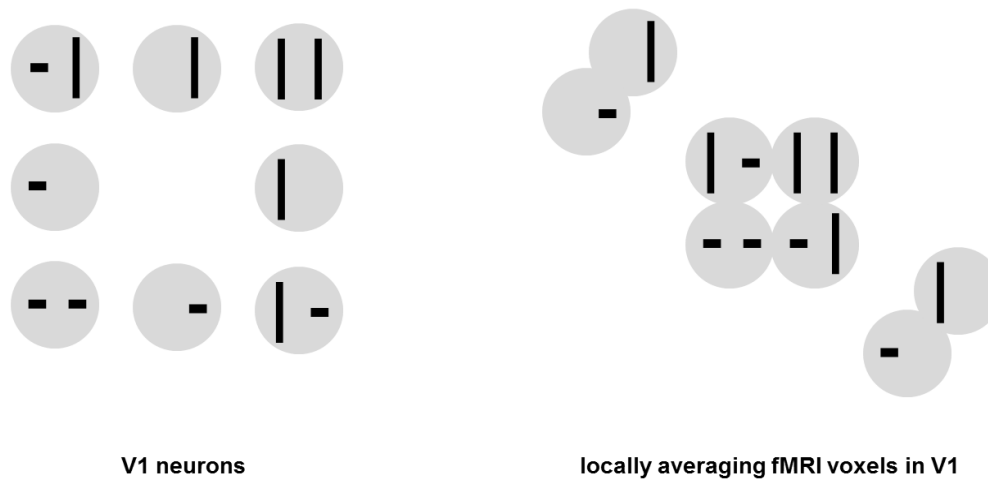

**V1 neurons**                    **locally averaging fMRI voxels in V1**

Figure 3 | **Distortion of the apparent representational geometry by locally averaging voxels.** This figure illustrates a worst-case scenario of how sampling with locally averaging fMRI voxels might distort the apparent representational dissimilarities. A set of 8 stimuli (gray discs with black symbols) is presented to a primate one by one, all at the same retinal location. In the V1 representation (left), bars of opposite orientation (horizontal, vertical) in the same location (left half of the disc, right half of the disc) drive separate sets of neurons. Bars in different locations within the disc also drive separate sets of neurons. The left panel shows the stimuli arranged in 2D, such that their distances on the page reflect their V1 representational dissimilarities. The stimulus pairs that are most distinct are stimulus pairs with opposite orientations in both locations (corner positions). When the V1 representation is sampled with locally averaging fMRI voxels (right panel), neurons preferring opposite orientations in a single location are pooled and the orientation information is attenuated (though likely not lost altogether, cf. Kamitani & Tong 2005; Kriegeskorte et al. 2010). We therefore expect two stimuli with bars in the same locations to elicit similar fMRI patterns, irrespective of the bars' orientation. As a result, the stimulus pairs that are most similar in the fMRI voxels are those that are most distinct in the neurons. The V1-neuron RDM is expected to be negatively correlated with the V1-voxel RDM. These observations hold for this particular stimulus set, not in general. The arrangements drawn here are approximate and designed to ensure visibility of all stimuli. (Qualitatively similar results obtain with MDS using Euclidean, cosine, or cityblock distance to measure representational dissimilarity and various MDS cost functions including metric stress, stress, and strain.)

A voxel sampling a little patch of V1 will average signals from neurons with very similar spatial receptive fields and selectivities for different orientations. We therefore expect that orientation-specific signals will be attenuated in the voxel patterns, whereas information about the overall spatial distribution of contrast across the image will be better preserved (Figs. 2-3). It is unlikely that a voxel samples all orientations (and similarly the entire range of selectivities for other properties like spatial frequency and colour) exactly equally (Kamitani & Tong 2005; Haynes & Rees 2005; Boynton 2005; Gardner 2009; Op de Beeck 2009; Kriegeskorte et al. 2010; Shmuel et al. 2010, Chaimow et al. 2011, Formisano &

Kriegeskorte 2012). To simulate an fMRI voxel, we assume the different features at the sampled location contribute to the signal with weights drawn uniformly from the interval [0,1].

## Measurement channels as samples from a population

An idea central to our method is that the measurement channels should be considered as samples from a population of possible measurement channels. All modalities of brain-activity measurement provide only subsamplings of the complex spatiotemporal dynamics of brain activity. Two-photon imaging enables us to measure a large proportion of the neurons in certain small animals, but it is limited in its temporal resolution. Electrode recordings are spatially and temporally precise, but limited to an almost vanishingly small subset of neurons. The voxels of fMRI, though strong in number and often covering the entire brain, average across tens or hundreds of thousands of neurons.

Not only do we take an informational subsample from neuronal activity, but the particular sample we take is determined by many factors beyond our control. We may target certain brain regions. However, the particular neurons we record from with electrodes or the particular sets of neurons sampled by our fMRI voxels are not under our control. Understanding the particular responses we happen upon is of limited interest. Rather the goal is to overcome the idiosyncrasies of a particular dataset and infer the underlying computational mechanism. To achieve this, we can view the measurement channels as a random sample from a population of possible measurement channels and target summary statistics of the population that can be robustly estimated from our particular sample and are rich enough to distinguish between candidate BCMs.

## Representational distances and the Johnson-Lindenstrauss Lemma

Representational distances, assembled in an RDM that contains a distance for each pair of stimuli, provide summary statistics with desirable properties. First, an RDM abstracts from the measurement channels and instead describes the relationships between representations. Second, an RDM provides a rich characterisation of a representation that is likely to discriminate functionally distinct brain representations. Information is lost in the summary. In the case of an RDM, what is lost is the information about which dimension of the representational space each measurement channel reflects. An RDM specifies an equivalence class of neuronal representations, all of which afford linear readout of the same properties of the stimuli (assuming that the noise is multinormal and the distance is the Mahalanobis distance).

When considering a human population such as the Dutch, we are very comfortable with the idea that we can rely on a random sample of, say, 500 Dutch people to estimate the average height of the Dutch population. If we measure representational distances as averaged squared activity differences, we likewise expect the average across a sufficiently large random sample of particular channels to be a good estimate of the average across the population of measurement channels. The RDM of average squared differences computed from our measurements will be a good approximation of the RDM we would obtain if we had the entire population of possible measurement channels.

The Johnson-Lindenstrauss Lemma (JL Lemma) states that distances between P points in an N-dimensional space are approximately preserved when the coordinates of the points are recoded along M < N new axes that are oriented randomly – as long as M > O(log P) [1] (Johnson & Lindenstrauss 1984). This means that for sufficiently large P, the minimally required M is proportional to log(P). Distances are approximately preserved when randomly weighted averages are taken (equivalent to projections onto randomly oriented axes) or dimensions are selected at random. Importantly, the number M > O(log P) of random samples required is independent of the dimensionality N of the original space.

This means that we expect a sufficiently large random sample of individual neurons from a population code to give us a good estimate of the RDM we would measure had we recorded the entire population. It also means that we expect a sufficient number of fMRI voxels to give us an RDM similar to that obtained from the neurons if the fMRI voxels sampled the neurons randomly (e.g. if neurons of different selectivity were randomly located in the measured region). This may explain why RDMs from cell recordings and fMRI can look surprisingly similar (Kriegeskorte et al. 2008b).

However, in general the neurons are not randomly located within a region, and so fMRI voxels do not take averages of random subsets of neurons. Voxels are, thus, not expected to give us RDMs that match those expected for neuronal recordings. For example, a voxel in V1 will average neurons with similar spatial receptive fields that are selective for different orientations. An RDM from fMRI is therefore expected to show weaker distinctions than an RDM from neurons between stimuli differing in local stimulus orientation, but similar distinctions between stimuli differing in the global distribution of contrast energy across the image (Fig. 3).

We can account for such distortions by modelling the measurements. To model locally-averaging measurement channels, we need to make predictions about the spatial layout of the BCM units, as provided, for example, by the visuotopic maps in deep convolutional neural networks. If an alternative spatial layout or measurement process appeared plausible, then this should be implemented as a generative model and included in the prior hypothesis space for Bayesian inference.

The RDM computed from a sufficient number of measurement channels is expected to be a good approximation of the RDM computed for the population of channels, of which the actual measurements can be considered a random sample. Note that this also implies that repeating the measurement and simulating the measurement (in either case taking a different random sample from the population of channels) should yield consistent RDMs.

## Probabilistic RSA

The JL Lemma is useful because it enables us to compute the RDM predicted by each BCM with an MM. For each BCM and setting of the MM parameters, there is a population of measurement channels, each sampling the representation at a different location. We can simulate a large sample from the population of measurement channels and compute an RDM prediction from that sample.

---

[1] Bachmann-Landau big-O notation

*Implementation of the fMRI measurement model.* Here the only MM parameter was the local-averaging range, which was specified as the full width at half maximum (FWHM) of a Gaussian kernel defining the local averaging. The FWHM was specified in units corresponding to grid steps for each convolutional map. For example, the first convolutional layer of AlexNet has maps that are 55 by 55 pixels large. Thus 1 unit corresponds to 1/55 of the visual angle spanned by the input image. We used a prior composed of a discrete set of 20 values equally spaced in the range [0, 16], thus covering a very wide interval of local-averaging ranges. In addition to Gaussian smoothing across retinotopic positions, the measurement channel simulation also pooled across different feature maps in each convolutional layer. After smoothing each feature map, each unit was assigned a weight drawn randomly from the interval [0,1]. For each retinotopic location, the weight vector across feature maps was scaled to sum to one. Then the weights were used to compute a weighted sum across feature maps. All weights used in this local averaging scheme were positive, reflecting the notion that fMRI voxels sample neuronal activity patterns by local integration (Kriegeskorte et al. 2009). Our scheme implements a particular model of the nature of fMRI measurements. Note that alternative models can easily be used in the same inferential framework. All that is required is a forward simulation of the measurement channels.

*Simulation of data for 12 subjects.* We simulated data from each of the five convolutional layers (BCMs) of AlexNet for 12 subjects. For each subject and BCM, we used a different value for the local averaging range and noise level, which were randomly chosen from the prior. Each simulated dataset comprised 500 simulated voxels, which were randomly sampled from a population of simulated voxels obtained by Gaussian smoothing (using the local-averaging range parameter to define the FWHM) and random weighting of feature maps using weights drawn uniformly from the unit interval. The MM parameters chosen in the data simulation, were not available to the statistical inference procedure, reflecting the fact that these are unknown parameters in the analysis of real data. The analysis had no information about which BCM generated the data (the target of the inference), the local-averaging range, the noise level, or the particular randomly chosen locations sampled by the simulated voxels.

*Crossvalidated Mahalanobis distance as measure of representational dissimilarity.* We used the crossvalidated Mahalanobis (crossnobis) distance, as the estimator of representational dissimilarity. Crossvalidated distance estimators are attractive because they are unbiased, reliable, and interpretable (Kriegeskorte et al. 2007; Nili et al. 2014; Allefeld & Haynes 2014; Walther et al. 2015). The Mahalanobis distance computed from noisy data provides a positively biased estimate of the Mahalanobis distances of the noisefree true patterns. For example, if the true distance is zero, the Mahalanobis distance will be positive whenever there is noise in the data. Classical RSA uses rank-correlation of RDMs in order to achieve robustness to biases and nonlinearities introduced by the way the representational dissimilarities are measured (Kriegeskorte et al. 2008a). The crossnobis distance is unbiased, that is its expected value matches the Mahalanobis distance. In particular, if the true distance is zero, the crossnobis distance estimate will be symmetrically distributed about zero. The interpretable zero point and undistorted (albeit noisy) reflection of the true distances may give us added power to distinguish among BCMs. Moreover, one of us recently derived a multinormal approximation of the distribution of crossnobis RDMs in closed form (Diedrichsen et al., 2016). This multinormal model enables us to evaluate the likelihood (probability of a crossnobis RDM estimate given a BCM and MM) and thus obviates the need to model the effect of noise by simulation or to rely on an approximate Bayesian computation (Sunnåker et al. 2013; Robert et al. 2011).

*Bayesian inference on brain-computational models.* For each subject, we performed Bayesian inference using Equation 1.

$$p(m|\mathbf{d}) \approx \frac{p(m) \cdot \sum_{i} p(\mathbf{d}|m, \boldsymbol{\theta}_i) \cdot p(\boldsymbol{\theta}_i)}{\sum_{j} p(m_j) \cdot \sum_{k} p(\mathbf{d}|m_j, \boldsymbol{\theta}_k) \cdot p(\boldsymbol{\theta}_k)} \quad (1)$$

The index $m$, for model, identifies the BCM. $p(m|\mathbf{d})$ is the posterior over models given the subject's data-based RDM estimate. The unique values of the RDM are stored in vector $\mathbf{d}$, which contains the crossnobis distance estimates in the lower triangular region of the condition-by-condition RDM. $p(m)$ is the prior over the BCMs, which we assume here to be uniform. $p(\mathbf{d}|m, \boldsymbol{\theta})$ is the likelihood, i.e. the probability of observed data RDM $\mathbf{d}$ given the BCM $m$ and the MM with parameter vector $\boldsymbol{\theta}$. In the present implementation, the MM has a single parameter that specifies the local-averaging range (FWHM). $p(\boldsymbol{\theta})$ is the prior over the MM parameter vector. The prior $p(\boldsymbol{\theta})$ is represented by a sample $\boldsymbol{\theta}_i$ of parameter values. We used an equally spaced set of 20 values ($i = 1, 2, ... 20$) of the local-averaging range, representing a uniform prior. We therefore set $p(\boldsymbol{\theta}_i) = 1/20$.

In order to evaluate the likelihood $p(\mathbf{d}|m, \boldsymbol{\theta})$, we used a recently derived multinormal model of the sampling distribution of crossnobis distance vectors $\mathbf{d}$ (Diedrichsen et al. 2016) in the space spanned by the $D = (K^2 - K)/2$ pairwise distances among the $K$ stimuli. First, the BCM $m$ is used to generate a response vector to each stimulus. Then the MM with parameter vector $\boldsymbol{\theta}$ is used to simulate a large sample from the population of measurement channels for each stimulus. A noisefree RDM $\mathbf{m}'(m, \boldsymbol{\theta})$ is computed for the stimulus set using the entire set of simulated measurement channels. We then fit a scaling factor $s$ that scales $\mathbf{m}'(m, \boldsymbol{\theta})$ so as to best fit the data RDM $\mathbf{d}$ in a least-squares sense:

$$s = \arg\min_{s} (\|s \cdot \mathbf{m}'(m, \boldsymbol{\theta}) - \mathbf{d}\|_2^2) \quad (2)$$

This estimate served our purposes here, but a better approach is to set $s$ by iteratively reweighted least squares, so as to maximize the likelihood in equation (4) below.

We refer to the scaled model RDM as:

$$\mathbf{m} = \mathbf{m}(m, \boldsymbol{\theta}, s) = s \cdot \mathbf{m}'(m, \boldsymbol{\theta}) \quad (3)$$

As derived in Diedrichsen et al. (2016), the likelihood $p(\mathbf{d}|m,\boldsymbol{\theta})$ is approximated by:

$$p(\mathbf{d}|m,\boldsymbol{\theta}) = p(\mathbf{d}|\mathbf{m}(m,\boldsymbol{\theta},s)) =$$
$$\exp\left[-\frac{D}{2}\log(2\pi) - \frac{1}{2}\log(|\mathbf{V}|) - \frac{1}{2}(\mathbf{d}-\mathbf{m})^T \mathbf{V}^{-1}(\mathbf{d}-\mathbf{m})\right] \quad (4)$$

This is a multinormal distribution centered on $\mathbf{m}$, whose shape is defined by stimulus-pair-by-stimulus-pair covariance matrix $\mathbf{V}$, with $D^2 = ((K^2 - K)/2)^2$ entries:

$$\mathbf{V} = \left[4\frac{\Delta \circ \Xi}{N_p} + 2\frac{\Xi \circ \Xi}{N_p(N_p - 1)}\right] \cdot \frac{\text{tr}(\boldsymbol{\Sigma}_R \cdot \boldsymbol{\Sigma}_R)}{P^2} \quad (5)$$

Here ∘ denotes element-by-element matrix multiplication. $\Delta$ is the stimulus-pair-by-stimulus-pair second-moments matrix of the activity differences among the true activity patterns. $\Xi$ is the stimulus-pair-by-stimulus-pair covariance matrix of estimated pattern differences across data partitions, which is a function of $\boldsymbol{\Sigma}_K$, the stimulus-by-stimulus covariance matrix of the measured response patterns. $N_p$ is the number of partitions of the data used to compute the crossvalidated (crossnobis) estimates of the Mahalanobis distances.

Note that $\mathbf{V}$ depends on both the model (via $\Delta$, a function of the assumed true activity patterns) and the data (via $\Xi$ and $\boldsymbol{\Sigma}_R$). $\Xi$ depends on $\boldsymbol{\Sigma}_K$, the stimulus-by-stimulus covariance matrix of the response-pattern estimates. $\boldsymbol{\Sigma}_R$ is the channel-by-channel covariance matrix of the residuals of the linear pattern estimation model.

In sum, for each BCM and each setting of the local averaging range, we simulated a noisefree RDM $\mathbf{m}'(m,\boldsymbol{\theta})$ for the set of 92 object images used in Kriegeskorte et al. (2008b). For a given single-subject data simulation, we computed RDM $\mathbf{d}$, $\boldsymbol{\Sigma}_K$, and $\boldsymbol{\Sigma}_R$. We used Equations (2) and (3) to fit the scale factor and compute $\mathbf{m}$. For each combination of a BCM $m$ and a MM parameter $\boldsymbol{\theta}$, this enabled us to compute $\mathbf{V}$ and the likelihood $p(\mathbf{d}|m,\boldsymbol{\theta})$ using Equations (4) and (5). For each BCM $m$, each of the 20 samples $\boldsymbol{\theta}_i$ represented an equal portion of prior probability mass $p(\boldsymbol{\theta}_i) = 1/20$. We therefore computed the marginal likelihood as the mean of the likelihoods $p(\mathbf{d}|m,\boldsymbol{\theta}_i)$ across $i = 1,2,...20$.

We assumed a uniform prior over BCMs. Thus, to obtain the posterior distribution (five probabilities here, one for each candidate BCM) for a given simulated data set for a single subject, we normalized the vector of marginal likelihoods to sum to 1. For each data-generating BCM, the group posterior over candidate BCMs was obtained by multiplying the single-subject posteriors and renormalising the vector of probability masses to sum to 1.

# Results

## The measurement process distorts the apparent representational geometry

We wanted to test whether the local averaging of simulated neurons (i.e. BCM units) by voxels would distort the apparent representational geometry and entail incorrect inferences when the measurement process was not accounted for. To this end, we analysed the simulated fMRI data for each BCM with classical frequentist RSA (Nili et al. 2014), using model RDMs computed from the BCMs without an MM (Fig. 4).

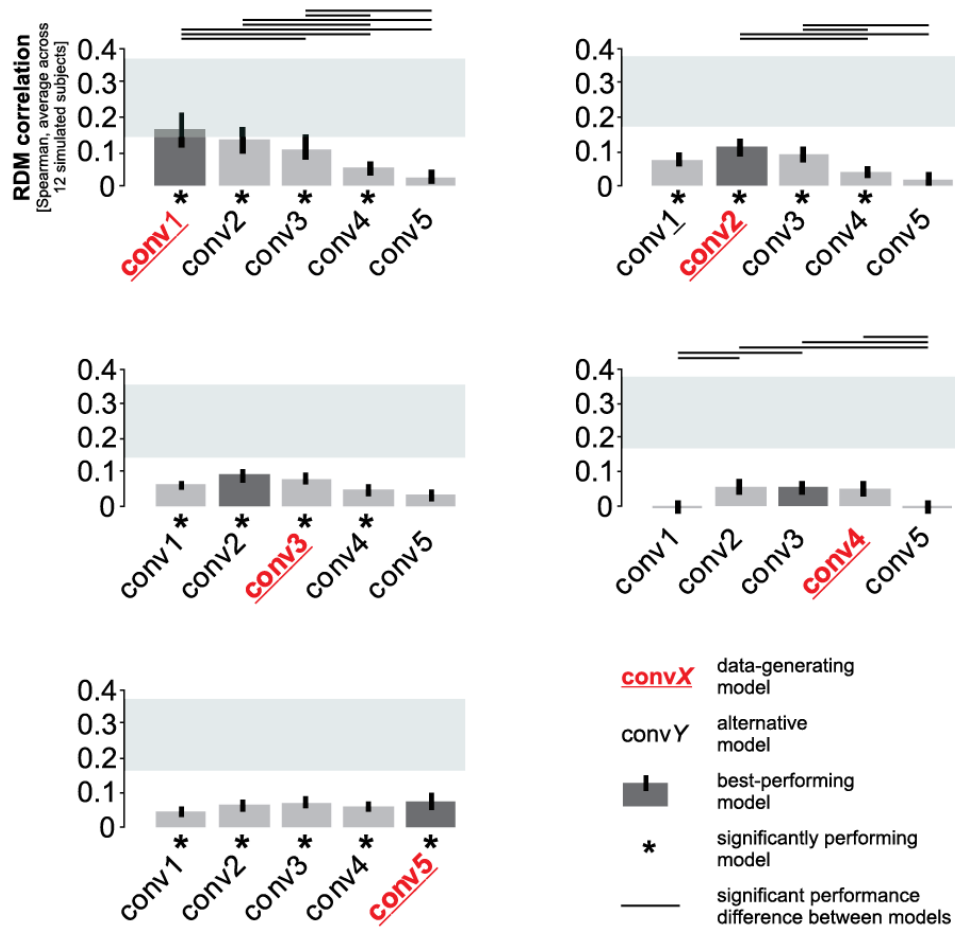

Figure 4 | **Classical RSA without modelling of the measurement process can fail to identify the data-generating BCM when measurements distort the representational geometry.** Frequentist RSA without a measurement model performed for simulated fMRI data from different layers of AlexNet. Voxels were simulated as local averages. Because the local averaging by the voxels is not accounted for in the analysis, the ground-truth data generating model (red text label) does not reach the noise ceiling (gray bar) for BCMs conv2-conv5. For two of five ground-truth BCMs (conv3 or conv4), the true model is not the best performing model. Group analysis for 12 simulated subjects. Stars indicate significant RDM correlations. Gray rectangles are noise ceilings, whose upper and lower edges indicate upper and lower bounds on the performance expected for the unknown true model given the noise in the data and the intersubject variability. Black lines above the noise ceilings indicate significant differences between models (subject-as-random-effect signed-rank tests, FDR<0.05). Note the substantial noise reflected in the low noise ceiling, approximately matching expectations for a small fMRI study. The same simulated data set was analysed with the proposed pRSA method without and with an MM in Figures 8 and 9.

Results clearly demonstrated the effects of the measurement on the RDMs. For all BCMs except convolutional layer 1, the data-generating BCM did not reach the noise ceiling, indicating that the model RDMs (not accounting for the measurements) failed to fully explain the data (simulated fMRI voxels that average locally). In three of five cases (conv1, conv 2, conv 5), the best performing BCM was still the ground-truth BCM that had generated the data. However, incorrect inferences did occur (conv3, conv 4), where another BCM achieved a higher RDM correlation than the data-generating BCM.

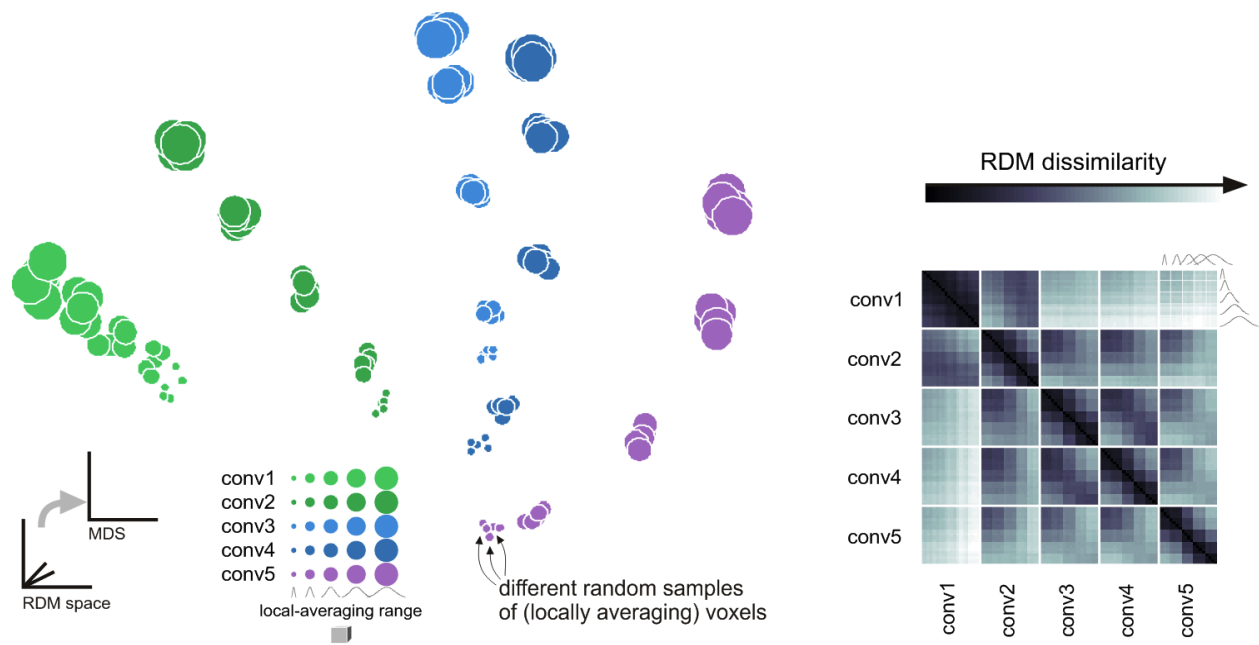

Figure 5 | **RDMs are robust to random sampling of locally averaging measurement channels, but strongly reflect the brain-computational model and the range of local averaging in measurement.** MDS (left) was used to visualise the pairwise differences among the RDMs (matrix of Euclidean distances between RDMs on the right). Each point corresponds to an RDM. The MDS arrangement is optimised for the 2D distances between points to best represent (in the sense of least squared deviations) the differences between RDMs. For clarity, the set of RDMs was restricted to 5 different local-averaging ranges (circle size). RDMs shown are for the no-noise condition, where random variability results only from the random sampling of 500 simulated voxels among the population of measurement channels. Clusters of overlapping points correspond to sets of 5 RDMs resulting from repeated sampling of the representation with a fresh set of 500 voxels placed at random locations in the BCM's representation. The difference between two RDMs was measured by the Euclidean distance between their dissimilarities. The MDS arrangement minimises the metric stress criterion.

The noise-level in the simulation was purposely set quite high, so as to pose a real challenge for inference. This is reflected in the low noise ceilings (gray rectangles in Fig. 4), which indicate the expected performance of the true model given the noise and intersubject variability (Nili et al. 2014). For this dataset, the true model never significantly outperformed all competing candidate models.

This shows that the RDM distortions caused by local averaging are not negligible in comparison to the RDM differences between BCMs. Measurement-related RDM distortions are large enough to mislead us about the data-generating BCM.

In order to visualise the effect of sampling BCM units with local averages on the RDMs, we performed multidimensional scaling (MDS) on the RDMs (Figs. 5, 6). Results confirm that local averaging strongly affects the RDM (Fig. 5). Two RDMs for the same BCM but different ranges of local averaging were often more different than two RDMs for different BCMs. In contrast to the worst-case scenario of Fig. 3, we used a set of 92 real-world object photos here. The strong effect of local averaging on the RDMs therefore cannot be dismissed as an idiosyncrasy of an artificial stimulus set.

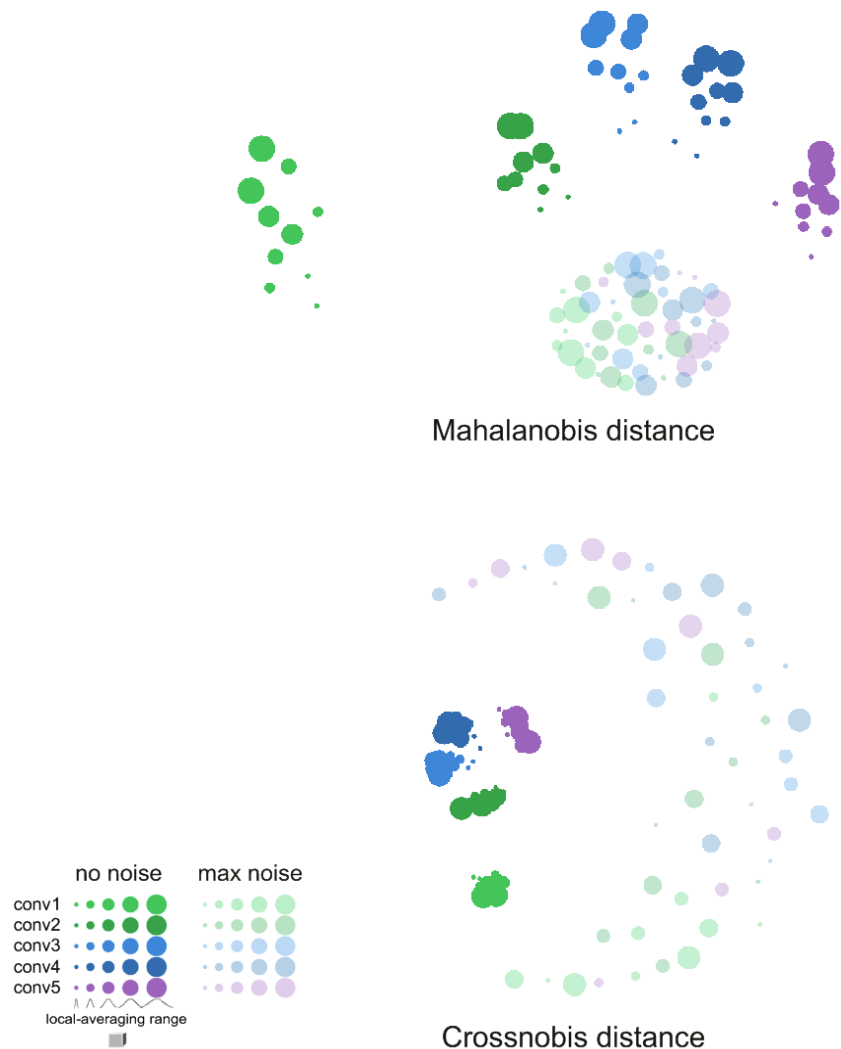

Figure 6 | **Noise mixes the RDM distributions associated with different BCMs.** For each BCM (conv1-5) and each MM (local-averaging kernel widths indicated by circle size), RDMs for two different noise levels are shown in MDS arrangements. In the no-noise condition, the only source of random variability in the RDMs is the random sampling of 500 locally averaging measurement channels. No noise is added to the channel responses. In the maximal noise condition, the amount of noise added is the upper limit of the noise levels used in simulating the single-subject data. Noise is differently reflected in Mahalanobis-distance RDMs (top) and crossnobis-distance RDMs (bottom). Noise nonlinearly distorts data-based Mahalonobis distances in comparison to the Mahalanobis distances among the true patterns. Noise creates a positive bias, which is strong for short distances and weak for long distances. High noise flattens Mahalanobis RDMs and makes them converge on a point in RDM space where all pairwise distances are equal and long. The crossnobis distance estimator, by contrast, is unbiased. Its expected values match the Mahalanobis distances among the true patterns. High noise, then, pushes RDMs in random directions and can mix the RDM distributions associated with different BCMs in the periphery. Conventions as in Figure 5. MDS based on Euclidean distances among RDMs, minimising metric stress.

Encouragingly, different BCMs are associated with separate distributions in RDM space. The local-averaging range moves the RDM along a one-dimensional manifold in RDM space (Fig. 5), but the manifolds for different BCMs do not intersect. Adding a high level of noise may lead to mixing of the RDM distributions (Fig. 6). Note, however, that apparent mixing of the distributions in the 2D MDS arrangement does not imply mixing in the high-dimensional RDM space. The remaining inferential analyses will address the degree to which the data-generating BCM can be inferred from a data RDM by modelling the predictive probability density over RDM space for each BCM with a prior of the local-averaging parameter of the MM.

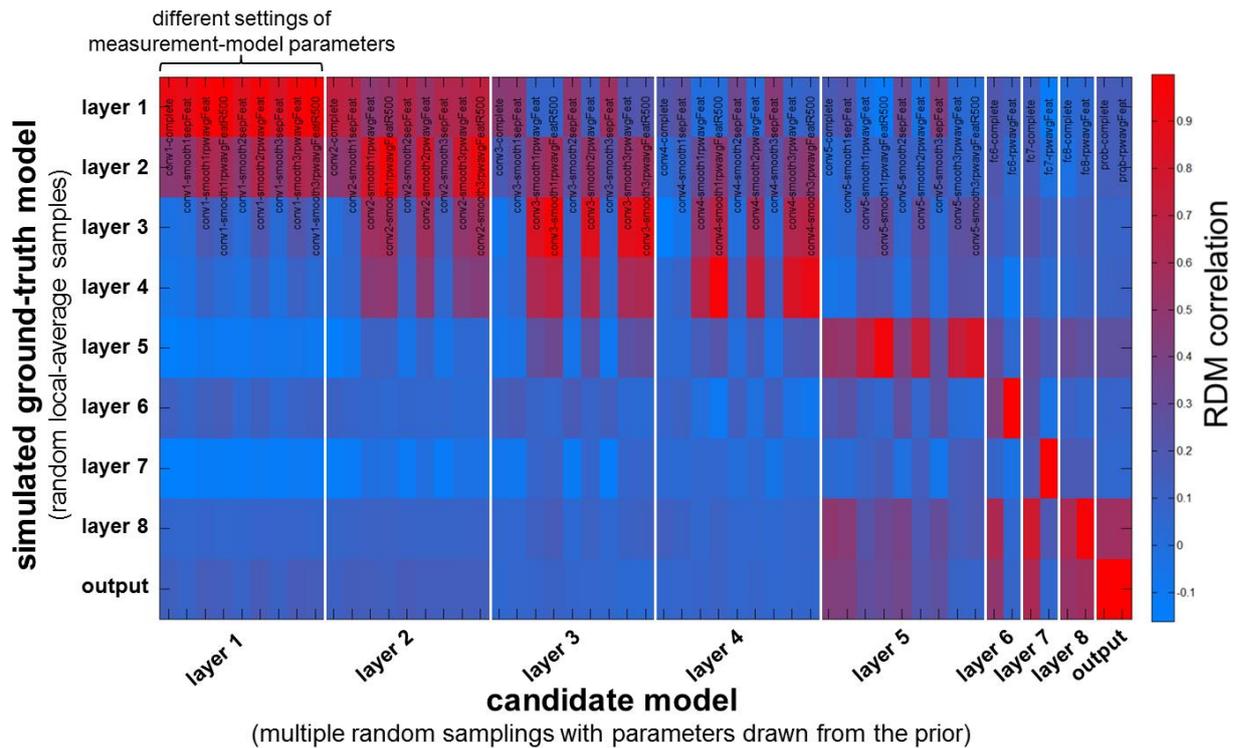

Figure 7 | **Inferring the data-generating computational model under uncertainty about how fMRI voxels downsample neuronal activity.** The matrix shows the RDM correlation between a simulated ground-truth RDM (vertical) and multiple simulated candidate model RDMs. The candidate RDMs are simulated without knowledge of the particular weights and locations used in simulating the ground-truth RDMs. Measurement model parameters strongly affect the RDM correlations (horizontal colour variation within blocks). However, for a given simulated ground-truth RDM, the candidate model RDM with the highest correlation is always from the correct BCM.

## Repeated simulations of the measurement yield consistent RDMs

Estimating the RDM for a given BCM and local averaging range repeatedly from different random samples of simulated measurement channels yields almost identical RDMs (clusters of overlapping circles in Fig. 5). This might be surprising in light of the fact that the RDMs are based on separate draws of 500 samples from the population of simulated fMRI voxels. However, this result is consistent with the JL Lemma. It means that the data-generating model can be inferred from RDM summary statistics. We need not model the idisyncrasies of the particular voxels we measured. Instead all we need is to model a random sample of locally-averaging measurements drawn from the same population of possible measurements as the actual measurements.

# Probabilistic RSA without a measurement model fails to accurately infer the data-generating BCM

This paper introduces both pRSA and the use of MMs. In order to understand the effect of each of these innovations, we first analysed the simulated data (same as analysed with classical RSA in Fig. 4) using pRSA without a measurement model. Each BCM was used to predict an RDM that was computed from all units of the convolutional layer in question, without taking local averaging samples. The noise was modelled using the multinormal model of the sampling distribution of crossnobis RDMs as explained in the Methods.

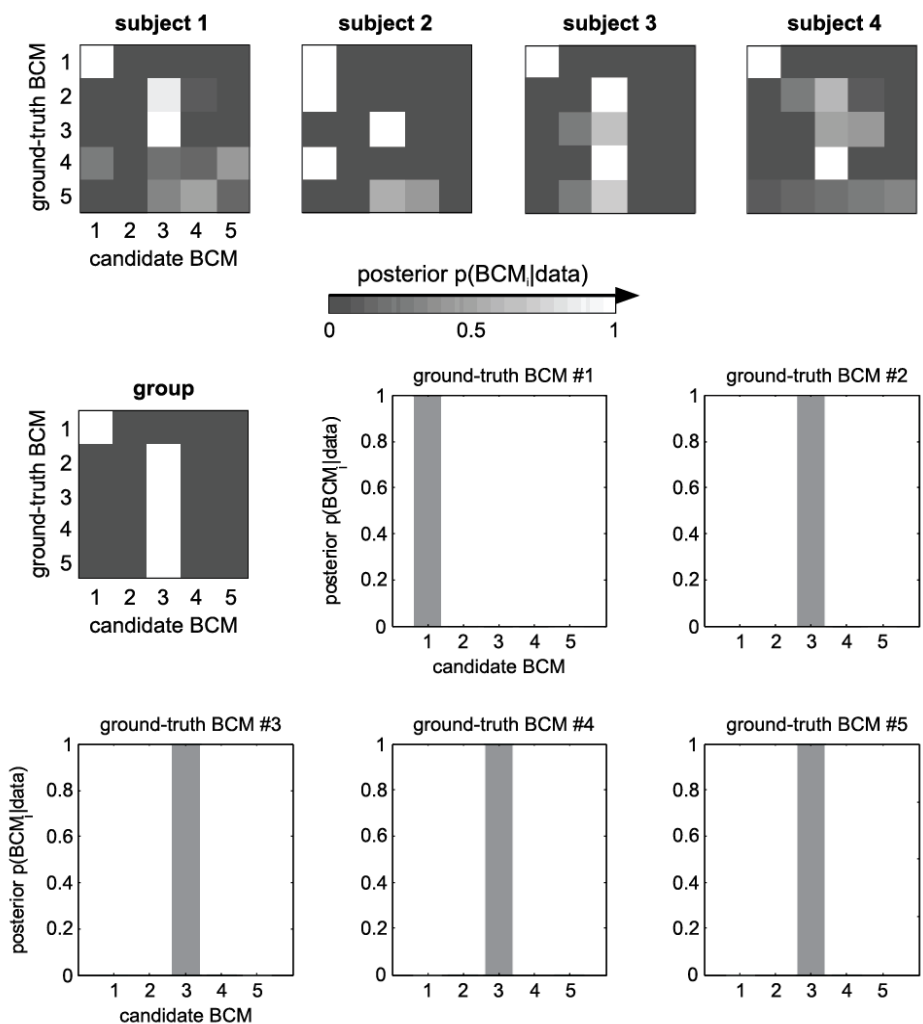

Figure 8 | **Without a measurement model, probabilistic RSA frequently fails to correctly infer the data-generating brain-computational model.** Probabilistic RSA was performed using the complete representation of each BCM without simulating the local averaging that gave rise to the data. Computing the posterior over BCMs for single simulated subjects (first four shown in top row) or for the group of 12 simulated subjects illustrates that ignoring the measurement process can yield incorrect results. The simulated data is identical to that analysed with classical RSA without a measurement model in Figure 4.

Although pRSA correctly recognised convolutional layers 1 and 3, assigning a posterior probability of nearly 1 to the data-generating model in both cases (Fig. 8), the analysis failed to recognise convolutional layers 2, 4 and 5. At the group level, the inference suggested that the data-generating model was convolutional layer 3 and the posterior probability assigned to layer 3 in each case again approached 1. This unsettling failure of probabilistic inference is explained by the fact that the inference is performed on the basis of incorrect assumptions. The analysis incorrectly assumed that one of the five layers must have generated the data and that the RDMs were computed from the original units without local averaging. The failure of inference highlights the need for modelling the measurement process.

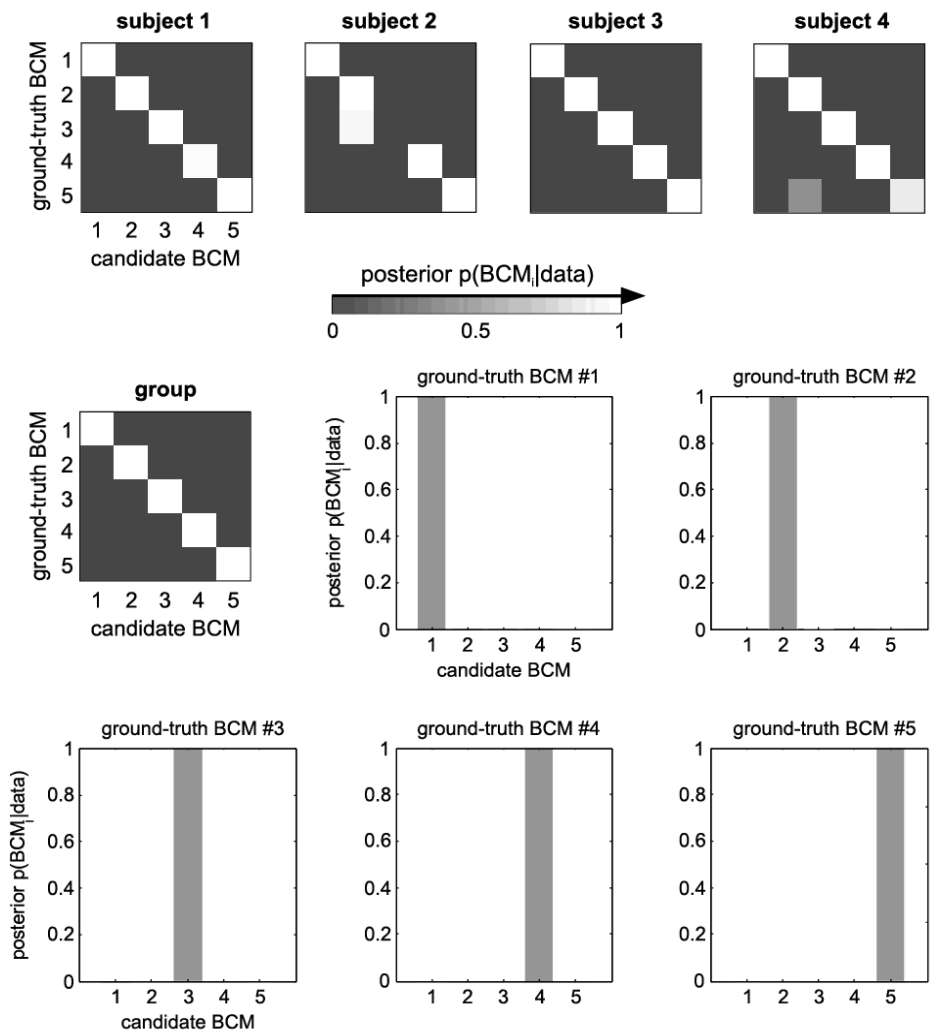

Figure 9 | **With a measurement model, probabilistic RSA correctly infers the data-generating brain-computational model.** When a measurement model with a broad prior over the local-averaging range was used, probabilistic RSA correctly identified the data-generating BCM in each case at the group level (12 simulated subjects), despite substantial noise (see noise ceiling in Figure 4) and distortions caused by the measurement process. Errors do occur at the single-subject level (e.g. for ground-truth BCM #3 in subject 2). The simulated data is identical to that analysed without a MM using classical RSA in Figure 4 and using pRSA in Figure 8.

## Probabilistic RSA using a measurement model accurately infers the data-generating BCM

Fig. 7 suggested that the effects of measurement can be accounted for by simulation. We performed pRSA with an MM parameterised by the local-averaging range (Fig. 9). We placed a broad uniform prior on the local-averaging range. The analysis was blind to the local-averaging range and noise level randomly chosen in simulating each subject's data. It had to take those uncertainties into account in the inference.

For each BCM and local-averaging range drawn from the prior, we predicted a Gaussian density in crossnobis RDM space, based on the multinormal model of the sampling distribution of crossnobis RDMs. The predictive probability density for a given BCM was thus a mixture of Gaussians. We marginalised the likelihood to obtain the model evidence and compute the posterior over the BCMs.

At the single-subject level, there were 60 data sets (each of 5 ground-truth BCMs for each of 12 simulated subjects). Of these 60 data sets, 2 were recognised incorrectly, i.e. the inference assigned an incorrect candidate model the highest posterior probability; 58 were recognised correctly. At the group level, all 5 BCMs were recognised correctly and assigned a posterior probability approaching 1.

These results suggest that pRSA with MMs works well when its assumptions are correct. It also suggests that the predictive distributions of the different BCMs do not overlap excessively in RDM space. As a result, the data-generating BCM can be accurately inferred from the data.

# Discussion

## Inference on complex BCMs with summary statistics

Realistic process models of brain computation require large numbers of computational elements, which may ultimately approach the number of neurons in the brain to be modelled. We need to test such models with massively multivariate brain-activity measurements. This poses a formidable correspondence problem: How do the units of a BCM affect each of the measurement channels? We have argued that for model comparison it is unnecessary to estimate a separate MM for each measurement channel. The idiosyncrasies of each channel in a given dataset are of no interest to us because they do not reflect the brain-computational mechanism. We showed that we can infer the data-generating BCM by predicting distributions of summary statistics of the stimulus-by-channel response matrix from each candidate BCM. We explored the RDM as a particular summary statistic. This led to the introduction of probabilistic RSA, in which the posterior over BCMs is estimated on the basis of the distributions of RDMs predicted by the BCMs.

## Modeling measurements as samples from a population of channels

Key insights were (1) that we can treat the measurements as a random sample from a population of potential measurements and (2) that we can model our knowledge about the way the measurements sample the computational units. A natural and flexible way to express what we know about the measurement process is by means of a forward simulation. Because we do not have precise knowledge of all determinants of the measurement process, we need a way to express our uncertainty. We can put priors on the unknown parameters of the MMs. The simulation draws the parameters of the MM from the prior, sampling the space of MMs considered possible. This enables us to account for the way a given computational process might be reflected in the brain-activity measurements.

In the method we implemented here, a BCM's predictive probability density function over crossnobis RDM space is a mixture of Gaussians, whose means trace a manifold parameterised by the MM parameters. The multinormal model of the sampling distribution of crossnobis RDMs (Diedrichsen et al. 2016) accounts for the effect of measurement noise and provides the Gaussian primitives from which the predictive density for each BCM is built.

We focussed on fMRI and assumed that the signals in fMRI voxels constitute local averages of computational units. A voxel reflects signals from a local region extending beyond its cuboid boundaries, because it reflects signals carried in through the local vasculature. We modelled this process as a local Gaussian-weighted average of the activity. Because we are uncertain about the cortical magnification (ratio of mm in the cortical map over degrees visual angle in the visual field) and about the range of signal integration beyond the boundaries of each voxel, we placed a prior on the width of the Gaussian filter, through which the measurements reflected the activity patterns across the computational units. Other modalities of brain-activity measurement will require different MMs. An advantage of our approach is that a forward simulation of how the measurements sample the neurons suffices for doing inference that takes the measurement process into account.

## Current limitations and future directions

The observations and ideas in this paper provide a starting point for pRSA. Likelihood-based inference promises greater sensitivity than classical rank-correlation-based RSA inference (Diedrichsen et al. 2016). However, additional development and validation will be needed before pRSA is ready for neuroscientific applications. Current limitations include: (1) The BCM-predicted RDM-distributional model is informed about the data via the scaling parameter $s$ and the covariance matrices $\Sigma_K$ and $\Sigma_R$. We need to determine whether this biases the inference and explore ways to model our uncertainty about these latent variables and marginalise across them. (2) It is unclear how our current implementation handles violations of the model assumptions. We need to implement model checking to infer whether all BCMs fail to explain the data. (3) The method has not yet been tested on real fMRI data. (4) Finally, a comprehensive quantitative comparison to frequentist RSA in terms of sensitivity (to BCM distinctions) and specificity (false-positives control) is still missing. Frequentist RSA correctly revealed that BCMs conv2-5 fell short of explaining the data when no MM was used in analysis (bars below noise ceiling). It also revealed its own inability to distinguish different BCMs (few significant differences). Measurement modelling could also be integrated into frequentist RSA. Ideally, our future method for inferring brain-computational mechanisms should combine the advantages of frequentist and Bayesian inference. Whatever this method turns out to be, it will need to take into account our knowledge and uncertainties about the measurement process.

## Acknowledgments

The authors thank Katherine Storrs for processing the images with the deep net and Johan Carlin for discussions about how to model fMRI measurements in the context of inference on brain-computational models. This research was funded by the UK Medical Research Council (Programme MC-A060-5PR20), by a European Research Council Starting Grant (ERC-2010-StG 261352) to N.K., and by a James S. McDonnell Scholar award to J.D.